# Estimation of Absolute States of Human Skeletal Muscle *via* Standard B-Mode Ultrasound Imaging and Deep Convolutional Neural Networks

Ryan J. Cunningham, Member, *IEEE*, and Ian D. Loram, *Member, IEEE*

*Abstract*— **Objective: To test automated *in vivo* estimation of active and passive skeletal muscle states using ultrasonic imaging. Background: Current technology (electromyography, dynamometry, shear wave imaging) provides no general, non-invasive method for online estimation of skeletal intramuscular states. Ultrasound (US) allows non-invasive imaging of muscle, yet current computational approaches have never achieved simultaneous extraction nor generalisation of independently varying, active and passive states. We use deep learning to investigate the generalizable content of 2D US muscle images. Method: US data synchronized with electromyography of the calf muscles, with measures of joint moment/angle were recorded from 32 healthy participants (7 female, ages: 27.5, 19-65). We extracted a region of interest of medial gastrocnemius and soleus using our prior developed accurate segmentation algorithm. From the segmented images, a deep convolutional neural network was trained to predict three absolute, drift-free, components of the neurobiomechanical state (activity, joint angle, joint moment) during experimentally designed, simultaneous, independent variation of passive (joint angle) and active (electromyography) inputs. Results: For all 32 held-out participants (16-fold cross-validation) the ankle joint angle, electromyography, and joint moment were estimated to accuracy 55±8%, 57±11%, and 46±9% respectively. Significance: With 2D US imaging, deep neural networks can encode in generalizable form, the activity-length-tension state relationship of muscle. Observation only, low power, 2D US imaging can provide a new category of technology for non-invasive estimation of neural output, length and tension in skeletal muscle. This proof of principle has value for personalised muscle diagnosis in pain, injury, neurological conditions, neuropathies, myopathies and ageing.**

## I. Introduction

There is a current unmet medical demand for personalized *in vivo* skeletal muscle analysis. Muscle related pain, injury and dysfunction represents an enormous socio-economic cost, including the cost of medical treatment, work absence, and long-term decreased ability to perform activities of daily living which exceeds that estimated for heart disease, cancer, or diabetes [1], [2]. This need arises in conditions of pain/injury (work-related injury, neck-back-leg pain and injury), arthritic conditions, neurological conditions (dystonia, motor neuron disease), myopathies (myositis), neuropathies (nerve injury, spinal cord injury) and changes associated with ageing (motor unit loss) [3], [4]. Work-related upper limb and neck musculoskeletal disorders are one of the most common occupational disorders around the world [5] and the cost of these in the EU has been estimated to be between 0.5% and 2% of gross national product [5]. Personalised diagnosis requires available, non-invasive, accurate, objective measurement of function and condition for skeletal muscles throughout the body [2]–[4], [6], [7].

The mechanical function of muscle is to deliver force. Muscle comprises muscle fibres embedded within a collagenous endomysia network [8]. This dynamic 3-dimensional structure, observable by US as shape and texture [7], transmits muscle force along the distributed curvilinear path between origin and insertion of each muscle [3], [4]. We hypothesize the dynamic state of skeletal muscle, is encoded by the 3D-collagenous structure, and is observable by 2D US images [3], [4].

Intrinsic muscle properties driven by two main independent inputs, active neural drive and length (origin-insertion distance), determine the dynamic state of muscle (Fig A Supp.M.). The state is termed "neurobiomechanical" because the state vector comprises one neural (activity) and two biomechanical (length, tension) components, defined here as (activity, joint angle, joint moment) [9]. Neural drive causes metabolically active contraction in muscle fibres. This internally generated pattern of tension contracts the internal collagenous structure, which shortens the muscle tissue and stretches the tendon tissue connecting muscle to bone. Joint angle reflects external forces (gravitational, contact, inertial) imposed on muscle. External force stretches the collagenous structure passively from outside and lengthens both the muscle and tendon. Because of the different active v passive force transmission patterns, we hypothesize the three components (activity, length, tension) are encoded instantaneously and independently within the collagenous structure and in a form generalizable between individuals.

Manuscript received xxx; revised xxxxx; accepted xxxx. Date of publication xxxxx; date of current version xxxx. This work was supported by Manchester Metropolitan University

Cunningham, and I. Loram are with the Cognitive Motor Function Group, Manchester Metropolitan University, M1 5GD,UK (email: Ryan.Cunningham@mmu.ac.uk, i.loram@mmu.ac.uk).

If correct, this hypothesis provides the basis for a new approach to acquire information from muscle. Particularly significant is the potential to measure neural output from deep muscles. Also significant is the potential to use the intrinsic collagen encoded activity-length-tension relationship to measure length and tension simultaneously using observation alone.

Current technology provides no general, non-invasive solution for measuring specific muscle states. Magnetic resonance (MR) allows low frame rate (< 10 Hz) imaging of musculoskeletal structures in inactive supine posture or limited movement [10]. Electromyography (EMG) subject to many well-known problems [11] can measure only the active component excluding the biomechanical (length, tension). Non-invasive, surface EMG is limited to superficial muscles, excluding general access to deep clinically important muscles in the neck, trunk and limbs. Intramuscular EMG can provide invasive measurement. Clinical neurophysiologists typically use needles and typically avoid deep muscles to prevent thoracic or spinal puncture. Dynamometry provides non-invasive measurement of joint moment and cannot provide muscle specific measurement.

US allows non-invasive real-time imaging of skeletal muscle to full anatomical depth (5-6 cm in the spine [7], up to 17 cm for the diaphragm [12]). Perturbation methods such as Supersonic Shear Wave Imaging and Shear Wave Elastography (SWE) [13], [14], induce a shear wave and measure its propagation through muscle [13]. Using multiple assumptions, the combination of a known stress with an observed strain pattern provides non-invasive estimates of regional stiffness within cross-sectional areas of specific muscles [14], [15]. There are limitations to SWE. Transmission power safety regulations limit depth of the adequate shear wave to 3-4 cm [16] which excludes the deepest muscles. SWE does not resolve active force (produced within the muscle by active motor unit firing) from passive force (caused by externally applied joint rotation or other external and internal factors). Correlations with measured force are subjective requiring calibration to person-specific maximum voluntary contraction (MVC). SWE theory limits acquisition to longitudinal scanning planes in line with the muscle fibres [17]. SWE has a maximum sampling rate of 1 Hz [14]. Standard frame-rate (25-100 Hz), full anatomical depth, b-mode imaging is clinically ubiquitous, non-invasive, low cost, and portable with minimal exclusion criteria. Observational analysis of strain only, is currently perceived as not revealing the tension state of skeletal muscle [16].

In summary, there is an unmet need for a non-invasive, estimation of skeletal muscle states. We ask the reader to view Video 1 (Supp. M.) which shows an US recording of the calf muscles undergoing simultaneous, independent change in active and passive input. From any single image, by comparison with any image selected as a baseline, could the reader estimate the instantaneous, absolute activity, joint angle and joint moment? This paper demonstrates proof of principle, that using deep learning (DL), and using standard observation-only, 2D US, three components of the dynamic neurobiomechanical state (activity, joint angle, joint moment) can be recovered directly from US images of muscle from people outside the training set.

## II. Context of Technical Contribution

Several authors have highlighted the nonlinear relationship between muscle image features (muscle thickness, length) and a singly varying external input such as EMG or joint angle e.g. [18]–[20]. With respect to skeletal muscle, research has focused on computational extraction of pre-defined, or partially defined, intuitive low-resolution features such as pennation angle, fascicle length, muscle shearing, fascicle curvature, muscle thickness, and cross-sectional area [18], [21]–[41]. Common limitations are lack of fully automated segmentation of muscles and features, manual initialization of analysis, confounding effect of non-muscle structures such as blood vessels, and cumulative drift arising from feature tracking methods [19], [22], [24], [30], [40]–[42]. Muscle is a complex 3-dimensional time varying structure in which features leave or enter the image plane: hence features are inherently impossible to track using pure feature tracking methods [43], [44]. Regulated tracking [24], [44] is closely related to feature engineering. Some presupposition about the information content is made and a technique is developed to measure and use that information to regulate spurious tracking points. However, consistent features suitable for regulation are generally lacking.

The intrinsic encoding, and estimation of more than one simultaneously varying component (activity, joint angle, joint moment), has never been demonstrated. Currently, no method has demonstrated automated robust (multiple held-out participants) generalization to new participants and no method extracts the complete state in general conditions where the inputs (activity, joint angle) vary independently.

The development [45]–[55] of DL provides a framework for encoding the content of US images in relation to measured data (EMG, angle, moment). DL is a technique for building artificial neural network (ANN) representations of data in a layer-wise fashion, where each layer models increasingly abstract/complex features of the data. DL facilitates modelling of complex features without *a priori* assumptions of the descriptive features. ANNs can learn nonlinear functions to map data (images) to labels (EMG, moment, joint angle). Even without many (or any) labels (which may often be the case with respect to deep muscles) features can be extracted using generative models such as restricted Boltzmann machines [45], [56], deep belief networks [57], deep (variational) auto encoders/autoassociators [47], [58]–[60], or more recently generative adversarial networks (GAN) [61]. Those features can either be analysed directly (using statistics or distance metrics), or re-mapped to relatively few labels. If large volumes of labelled data exist, a CNN can be trained directly on the data to predict the labels, which can be continuous or discrete. CNNs work well for understanding the content of static images [62] or speech [63], and more recently deep residual networks (ResNet) have surpassed human-level performance [64] on an the ImageNet image recognition competition. CNNs have also demonstrated the ability to track local motion [65], which means that unlike standard feature tracking, a CNN can measure the dynamic state (2 temporally different frames) of local features, while simultaneously having access to the static state (or pose) information. Use of historical states, perhaps with recurrent or long short-term memory networks, can also be valuable. For this investigation of a new hypothesis (that the instantaneous collagen structure encodes states) we avoided temporal ANN models since they cloud the issue, are comparatively difficult and time-consuming to train and would complicate generalization to different US acquisition rates (e.g. ultrafast US > 1000Hz *vs* standard 25-100Hz).

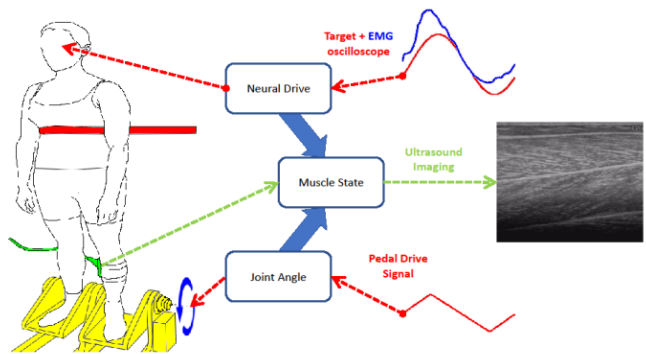

**Figure 1. Experimental setup.** The participant stood upright on a foot pedal system (yellow), while strapped (red) at the chest to a backboard and observed an oscilloscope at eye level. An US probe (green) was attached to the left calf to image the gastrocnemius medialis (GM) and soleus (SO) muscles (right: grayscale). A wireless EMG sensor was attached to GM and to SO at standard locations (http://www.seniam.org/). By contracting their calf muscles, the participant matched the GM EMG feedback signal (blue) to the target signal (red) presented on the oscilloscope. A pedal signal (red) rotated the pedals and ankle joint angle at the rotational axis (blue arrows).

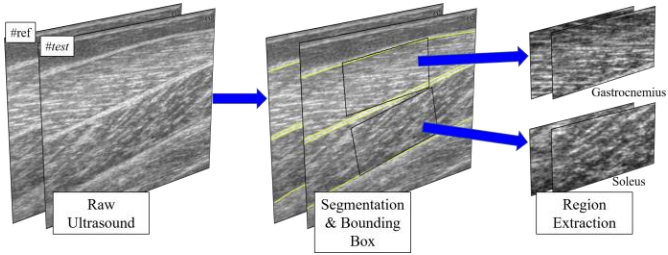

**Figure 2. Image segmentation and region extraction pipeline.** The fully automatic region extraction process occurs prior to neural network training and testing. **Left**: a pair of US images is shown, where the #ref image represents the *reference* image; the *reference* image accompanies a #test image as part of a single input to the neural network. **Middle**: yellow lines show the automatic segmentation of individual muscles, from which bounding boxes (black rectangles) are positioned on the centroid oriented orthogonal to the main axis of each muscle. **Right:** a $128 \times 256$-pixel region is extracted from the raw images. For each muscle, the label vector (muscle EMG, joint angle, joint moment) equals the values temporally aligned to the *test* frame, subtracted from the values temporally aligned to the *reference* frame.

We use deep learning (CNNs) to map individual frames with a contextual reference frame (prior), to absolute (drift-free) states measured by other means (ankle angle, muscle EMG). This work is not a study of different CNN architectures. This work tests, for the first time, whether or not the three components of the absolute neurobiomechanical state can be estimated using DL. This work is informed by our own prior investigation [66], combined with developments in the field [64], [67].

## III. METHODS

### A. Experimental Design and overview of methods

We test the hypothesis that US images alone contain the information required to model the state of muscle and to resolve that state into the two independent inputs which created it. We identify the two main independent inputs as muscle activity and joint angle. We select two muscles, gastrocnemius medialis (GM) and soleus (SO), for which both inputs can be manipulated and measured to provide ground truth. We design apparatus and a protocol, which allows us to vary each input independently and simultaneously, to create a dataset of muscle US images populating the space of possible neurobiomechanical states. We investigate a form of supervised learning (CNNs) for their potential to learn, with minimal over fitting, the temporal variation of biomechanical states and we use 16-fold cross-validation providing genuine held-out test results for all 32 of our participants to test our hypothesis.

### B. Data acquisition

Thirty-two healthy participants (7 female, ages 19-65 mean 27.5) stood upright on a programmable/controllable foot pedal system while strapped at the hip to a backboard (Fig. 1).

Joint angle refers to the angle between the foot and the shin. The participant was restrained, maintaining a straight leg and standing flatfoot on the pedal system. Joint angle is measured by rotation of the pedals from horizontal. The calf muscles deliver force through the Achilles tendon. This force rotates the foot relative to the shin. Joint moment is the rotational effect of the combined muscle forces acting around the joint axis of rotation. For the range of motion studied, ankle moment is approximately proportional to the summed calf muscle force. We measured ankle moment using a gauge built into the foot pedals.

Electromyography (EMG) is the electrical activity arising from the active contraction of the muscle fibres. Surface rather than intramuscular electrodes provide the best global measure of muscle activity [68], [69] and so this electrical activity is recorded from the skin surface above the muscle using electrodes (Trigno, Delsys Inc., USA). We recorded EMG from the GM muscle belly, EMG from SO at its medial superficial location, ankle joint angle and ankle joint moment, all at $1000Hz$.

Using an US scanner (Aloka Prosound SSD 4000+, probe 7.5 MHz, width x depth 5.9 x 5.5 cm), we imaged GM and SO simultaneously in their longitudinal fascicle plane. The imaging location, angle and depth was chosen to include, within plane fascicle collagen content of both muscles, but optimised for GM. The probe was strapped to the participant to maintain constant location during movement [19]. US was recorded at $25Hz$ recording using a frame grabber (DT 3120, Data Translation). We used Simulink (Matlab, R2013a, The MathWorks Inc., Natick, MA) to interface with the lab equipment (pedal system and EMG). For video synchronization a hardware trigger was used to initiate at the start of each trial.

### C. Tasks

Three distinct tasks were designed to explore the state-function space of muscle:

*1) Isometric*

The pedal system was fixed at a neutral angle (flat feet), and participants observed an analogue oscilloscope. On the oscilloscope, we displayed, side by side, a dot representing the amplitude of their filtered GM EMG signal, and a dot representing the amplitude of a fabricated (target) signal (see section III. B.). Participants were asked to contract their calf muscles by pushing down their toes in order to match their EMG with the target

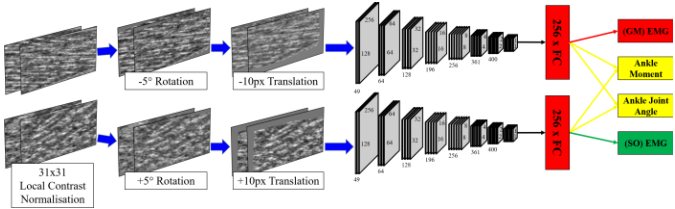

**Figure 3. Data augmentation and CNN pipeline.**
**Online data augmentation**. First, we normalise contrast by removing mean and variance, sampling from patches of ± 15 pixels. Then we sample a random rotation parameter (± 5 °) per muscle and rotate each pair of muscle images independently. Finally, we sample a random translation parameter (± 10 pixels) per muscle and translate each pair of muscle images independently.
**CNN pipeline**: two CNNs with shared parameters and a dense layer (FC) encode each muscle separately. In the final layer, 4 linear units connect to the FC layers of each CNN: the ankle moment and ankle joint angle units are fully connected to the units of both CNNs, whilst the GM EMG unit is only connected to the CNN which receives input from the GM image. Also, the SO EMG unit is only connected to the CNN which receives input from the SO image.

signal, while simultaneously keeping their foot in full contact with the static pedals.

*2) Passive*

Participants observed an analogue oscilloscope. On the oscilloscope, we displayed, side by side, a dot representing the amplitude of their filtered GM EMG signal. Participants were asked to monitor and minimize any EMG activity by relaxing their muscles. The pedal system was driven using a fabricated signal (c.f. III. B.). Participants were asked to allow their ankle to rotate and keep their feet in full contact with the moving pedals.

*3) Combined*

The pedal system was fixed at a neutral angle (flat feet), and participants observed an analogue oscilloscope. On the oscilloscope, we displayed, side by side, a dot representing the amplitude of their filtered GM EMG signal, and a dot representing the amplitude of a fabricated (target) signal (III. B.). The pedal system was simultaneously driven using a different fabricated signal (III. B.). Participants were asked to contract their calf muscles by pushing down their toes in order to match their EMG with the target signal, while simultaneously keeping their foot in full contact with the static pedals.

Trials were 190 seconds in length which consisted of 10 seconds of neutral standing (i.e. no signals were used to move the pedals or the dot on the screen), followed by 180 seconds of trial. Data ranges are shown in Table 2.

### D. Designing the labels

Two signals (active contraction, passive joint rotation) were designed to manipulate the two independent muscle inputs. Both signals were derived from the following bases:

*Active contraction* (target dot on the screen to guide calf muscle contraction). 1) For the first 10 seconds signal *a* was used, and every 10 seconds thereafter we alternated between signals *a* and *b*. 2.) After 30 seconds signal *c* was used, and every 30 seconds thereafter, either signal *a* or *b* was used depending on the first rule.

*Passive joint rotation* (pedal angle): 1) For the first 20 seconds signal *a* was used, and every 20 seconds thereafter we alternated between signals *a* and *b*. 2). After 60 seconds signal *c* was used, and every 60 seconds thereafter, either signal *a* or *b* was used depending on the first rule. The signals were designed to produce transient correlations, de-correlations, and anti-correlations to maximize exploration of the muscle input state space. The correlation of the two signals was $r = 0.33, p = 0$ (Pearson), and $r = 0.34, p = 0$ (Spearman).

### E. Segmentation and region extraction

$$a = \sin\left(0.4t\pi - \frac{\pi}{2}\right),$$
$$b = \sin\left(0.5t\pi - \frac{\pi}{2}\right), \quad (1)$$
$$c = \sin\left(\sin\left(t\frac{\pi}{30} - \frac{\pi}{2}\right)30\pi - \frac{\pi}{2}\right).$$

To map muscle specific EMG to muscle specific image we extracted regions of superficial (GM) and deep (SO) muscle tissue (Fig. 2). Analysis of restricted, standardized regions enabled us to maximize the spatial resolution while reducing the computational dimensions and complexity. First, an expert annotated the internal boundaries of the medial GM and SO muscles in 500 randomly selected images of which 100 were selected randomly for testing. After interpolating the annotations to a standard 80-point vector, a principal component model was constructed from the remaining 400 images. An ASM [70], constructed from just 10 principal components ($> 99\%$), was used to guide a heuristic search with a large profile search range ($\pm 30$ pixels about the contour). No initial segmentation was used; the increased profile range was an ample aid to match distant contours. The search was conducted at full resolution $\pm 10$ pixels about each contour point. For more detail see [7]. The entire dataset ($> 400,000$ images) was segmented.

To standardize the image input, we extracted a rectilinear region ($x \times y = 256 \times 128\ pixels \approx 29.4 \times 14.6\ mm$, Fig. 2) about the centroid of each muscle, orthogonal to the main axis of the muscle. The main axis was calculated as linear least square fit to mean segmentation over the whole trial sequence. This region captures the muscle tissue rather than the tendon which connects muscle to bone.

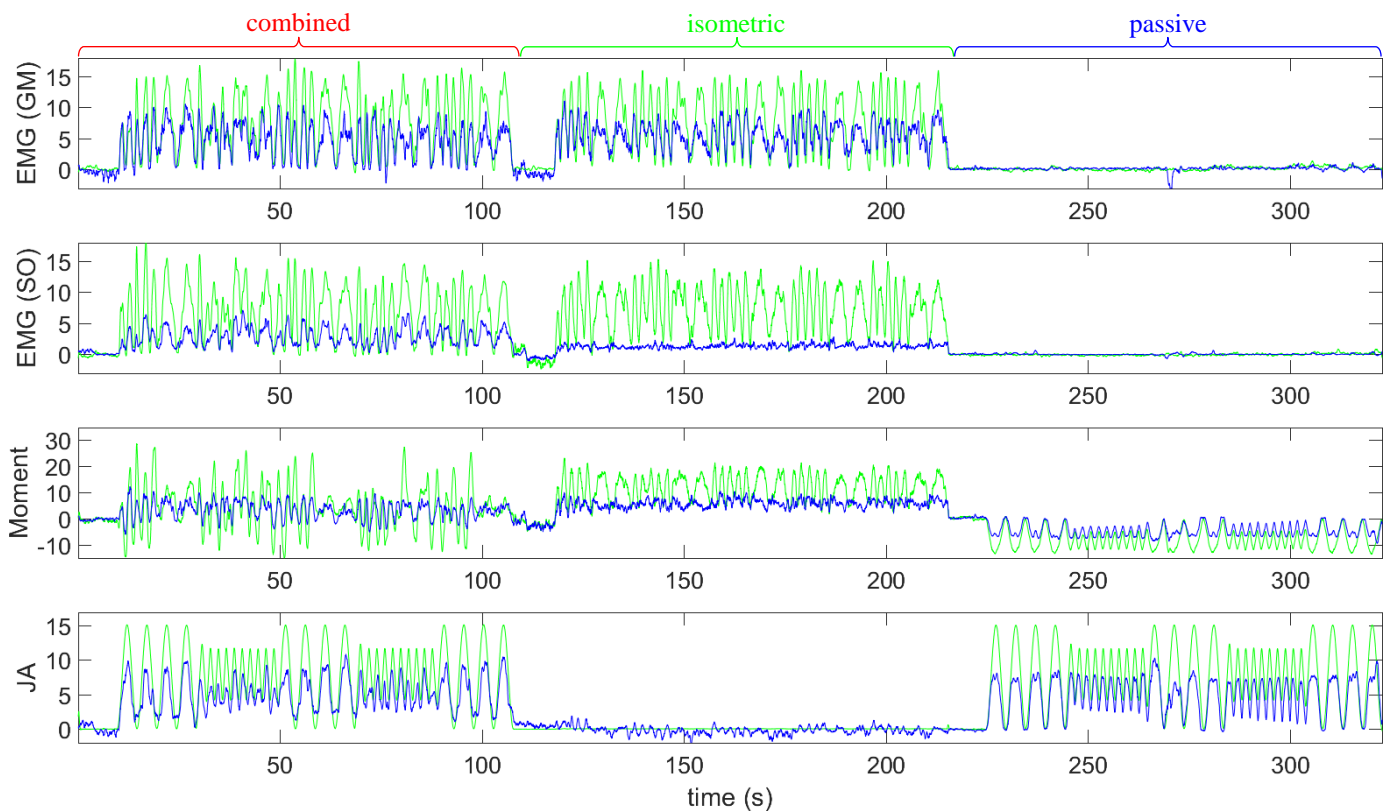

**Figure 4. Representative example of neural network performance.** Each panel compares the neural network output (blue) with the labels (green) for a single participant over all 3 trial conditions (*combined*: left, *isometric*: middle, and *passive*: right). Units are mV, Nm and degrees for EMG, Moment and Joint Angle.

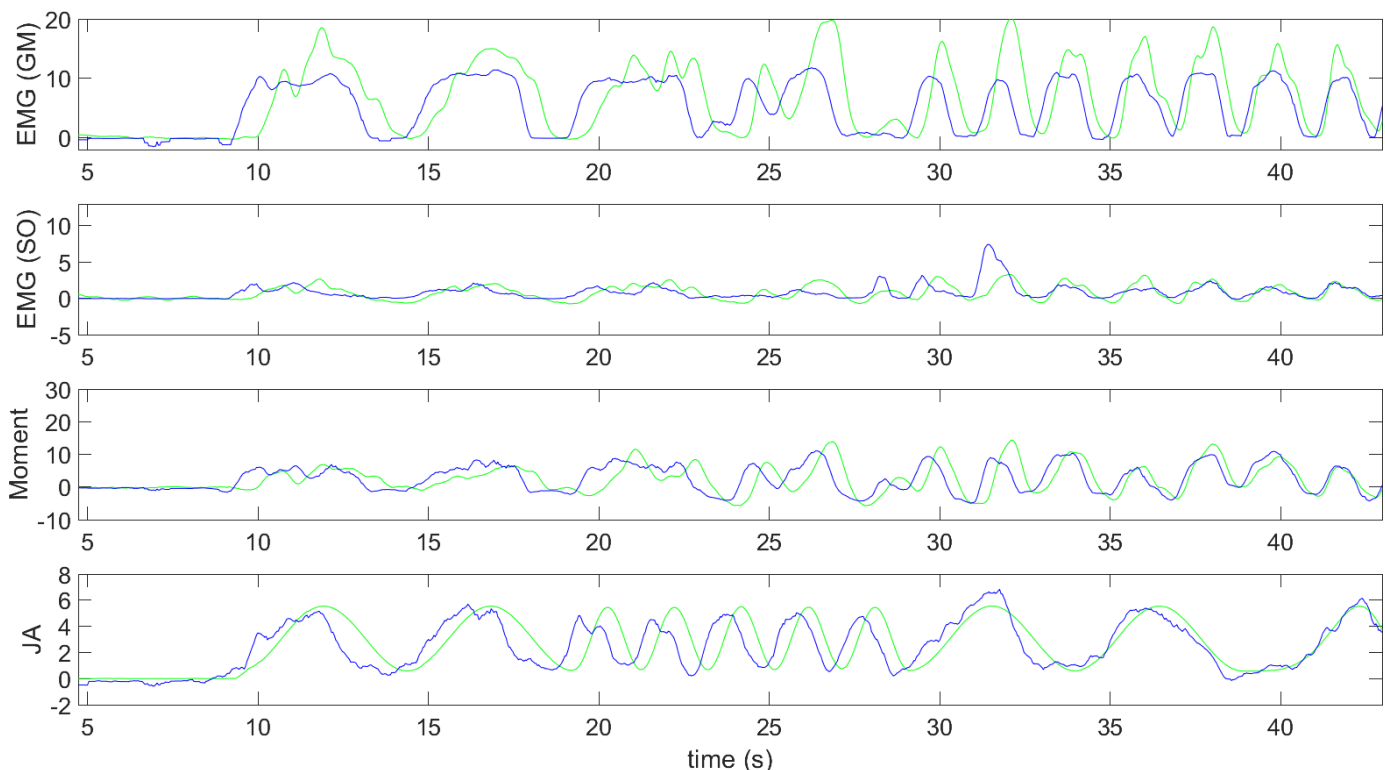

**Figure 5. Zoom portion of a representative participant.** The zoom shows ~40 seconds of contiguous data from a single participant during a trial constructing combined, independent modulations of ankle joint angle, and GM/SO EMG. The graphic illustrates how the neural network has separated and modelled the 4 independent signals in a representative case. The participant differs from Fig. 4. Units are mV, Nm and degrees for EMG, Moment and Joint Angle

**Table 1. Summary of neural network test results.** This summarizes the results presented in table A (supplementary material). We present mean, standard deviation, median, minimum and maximum values for Symmetric Mean Absolute Accuracy (100-SMAPE) and Mean Absolute Error (MAE), calculated over all 32 participants, for each of the 4 labels: EMG (GM/SO), ankle joint moment, and ankle joint angle (JA). Samples are images per participant: the sum reports all images tested from the complete data set of 32 participants.

| Metric | Samples | Symmetric Mean Absolute Accuracy (100 - SMAPE) | | | | Mean Absolute Error (MAE) | | | |
|---|---|---|---|---|---|---|---|---|---|
| | | EMG | | | | EMG | | | |
| | | GM (%) | SO (%) | Moment (%) | JA (%) | GM (mV) | SO (mV) | Moment (Nm) | JA (°) |
| Mean | 12,594 | 56.90 | 45.91 | 46.97 | 54.57 | 3.06 | 2.61 | 6.11 | 2.50 |
| Standard Deviation | 2,613 | 11.02 | 11.06 | 8.87 | 7.94 | 1.02 | 1.57 | 3.04 | 1.28 |
| Median | 13,944 | 59.37 | 45.33 | 47.60 | 53.82 | 2.88 | 2.18 | 5.07 | 2.19 |
| Min | 8,058 | 35.61 | 24.26 | 27.49 | 43.33 | 0.59 | 0.55 | 2.58 | 1.06 |
| Max | 14,971 | 72.45 | 69.06 | 61.54 | 67.30 | 5.84 | 7.56 | 15.97 | 6.17 |
| Sum | 403,023 | - | - | - | - | - | - | - | - |

**Table 2. Label statistics.** Table shows the distribution of labels recorded over all 32 participants. Negative joint angle represents dorsiflexion (decrease in angle between foot and shin referenced from flatfoot i.e. 90 °). Positive angle represents plantarflexion (increase in angle between foot and shin referenced from 90 °).

| | EMG | | Moment (Nm) | JA (°) |
|---|---|---|---|---|
| | GM (mV) | SO (mV) | | |
| Mean | 5.62 | 4.67 | 16.92 | 1.24 |
| Standard Deviation | 5.32 | 4.70 | 11.30 | 3.51 |
| Median | 2.74 | 2.87 | 15.23 | 0.66 |
| Min | 0.0026 | 0.0021 | -14.77 | -9.32 |
| Max | 50.21 | 89.58 | 94.35 | 13.79 |

### F. Neural network architecture

Our primary concern was to choose an architecture for which the model was large enough to minimize adequately the training error. The second main concern was to maximise generalization and minimise computation. Using our previous experience [66], our strategy was to train a large model, with state of the art regularization (dropout) in multiple layers, while evaluating performance on held out validation data. Unlike our original study [66] we additionally address the deep muscle (SO), and this decision inspired a CNN architecture wherein a CNN model is created and applied per muscle, and the weights were shared between models as a regulariser (Fig. 3). We are not the first to use this type of architecture [71], though the application if it is novel.

The learning objectives are different for each of the muscle-specific CNN models. The muscle-specific EMG should be predicted from the relevant muscle to ensure the estimation of muscle activity came from the target muscle. However, the ankle angle and ankle moment should be estimated from the states of both muscles combined. To meet this objective, information/gradient flow was gated using a binary mask for the relevant learning objectives (Figs 2 & 3). Data augmentation was implemented online at the input to each CNN (Fig. 3).

### G. Online data augmentation

To aid convergence as well as generalization, local contrast normalization (LCN) was applied to each image via a GPU with a local field of 31 square pixels (Fig 3). During training, to help prevent overfitting and account for (intra/inter)-participant variation in muscle region extraction, linear transformations (rotation and translation) were applied to the reference image and the target image per muscle. The same transformation was applied between reference and target images, but not between muscles (i.e. each muscle had its own transformation). Rotations and translations were randomly sampled from a uniform distribution between -5° and 5° and -16 and 16 horizontal/vertical pixels, respectively. Transformations were carried out on a GPU using linear interpolation with no extrapolation, where zeros filled the extra pixels (Fig 3).

### H. Training and cross validation

The input to the neural network was a pair of images; a reference image (frame #1), and a target image (any frame within a trial), and the labels. The labels for each input were the test frame aligned EMG, ankle angle and ankle moment minus the reference frame aligned EMG, angle and moment. Labels were normalised to unit standard deviation. The learning objective was to predict the difference in the independent states of the muscles between two images. To reduce bias in the input channel of the network corresponding to the reference image, we doubled the training set by swapping the reference and target frames. We randomly sampling reference and target frames to increase further the size of the training set. The final training set contained over 1 million pairs of images.

To train our models we minimized the mean absolute error (MAE) between the model output and the normalised labels (EMG, joint moment, joint angle) using Adaptive Moment Estimation (ADAM: [72]) with $alpha = 0.999$ and $beta$ 0.9 and a learning rate of $5e-5$. To prevent saturating units [66], Exponential Linear Units were used in all layers except the output layer which was linear. Prior to training, all biases were initialized to 0, and all weights were initialized using,

$$var(w) = \frac{2}{fan\_in}.$$

where $w$ is the normal distributed weight vector of a single unit/node and *fan_in* is the size of the input vector to that unit.

The train, test and validation errors were measured periodically during training to allow selection of optimal models using test and validation errors. We used 16-fold cross-validation: for each fold unique combinations of validation and test data were used to assess performance of the model within the fold. Within each fold a test set of one held out participant (approx. 12,500 samples) and a validation set of one held out participant (approx. 12,500 samples) was created (where none of the test/validation participants were used in training). For each fold, the validation set was used to choose the optimal test model and the test set was used to select the optimal validation model. This process yielded 16 unique neural networks with genuine held-out results for all 32 participants. To regularize our models, we used a dropout scheme similar to [67], where dropout was applied to every layer with larger dropout rates in layers closer to the output of the model. That dropout strategy circumvented the need to try variations of dropout, requiring repeated training and model evaluation as in [66]. As additional regularization and detection of convergence, early stopping was used where the model with the lowest test/validation error was taken after both test and validation errors did not decrease for more than 8 error evaluations.

Errors are reported using mean absolute error (MAE) for each signal. To report accuracy of estimate $Y$ in the context of time varying signal $y$, we use,

$$\text{Accuracy} = 100 - \text{SMAPE}$$

where SMAPE (Symmetric mean absolute percentage error) is:

$$SMAPE = \frac{100\%}{n}\sum_{i=1}^{n}\frac{|Y_i - y_i|}{(|Y_i|+|y_i|)/2}.$$

*I. Software and Tools*

All ANN and segmentation software was developed from first principles by the 1$^{st}$ author using C/C++ and CUDA-C (NVidia Corporation, California). No libraries other than std CUDA libraries (runtime version 8.0 cuda.h, cuda_runtime.h, curand.h, curand_kernel.h, cuda_occupancy.h, and device_functions.h), the C++ 11 std library and OpenMP were used.

IV. RESULTS

*A. Segmentation and region extraction*

Our requirement for segmentation is accuracy within the dataset rather than generalization. Cropping well within the muscle boundary and averaging over the sequence requires millimetre rather than sub-millimetre accuracy. For the 100 randomly selected test images, segmentation agreed with the manual annotations to $0.3 \ mm^2$ (~99% IoU) and segmented at *approx.* 10 images per second.

*B. Representative Neural network output*

The CNN estimates the neurobiomechanical state of a muscle (EMG, moment, angle) for each frame independently of all other frames (except the reference). The extended representative sequence from one participant shown in Fig. 4 illustrates there is no drift in the estimated state components.

Increased activity shortens muscle, stretches the tendon and increases tension, whereas positive joint rotation (plantar flexion) shortens the muscle passively and decreases muscle (and tendon) tension. Activity and positive joint rotation both reduce strain in (shorten) the collagen structure but have opposite effects on tension in the same collagen structure. A key question is whether the CNN can resolve these independent active and passive changes in muscle.

Consider Fig 4, here we describe qualitatively accuracy of the estimate in terms of the temporal pattern of states between frames, the local timing of the estimated v actual pattern, the scale and the bias of the local pattern. Fig. 4 shows the CNN captures the pattern of GM EMG during isometric and combined conditions and correctly shows no activity during passive joint rotation. The CNN estimates change in GM activity independently from joint rotation in isometric, passive and combined conditions. However, the scale is too small giving an estimated GM EMG approximately 50% of the true signal. For the deep muscle SO, the CNN distinguishes activity (combined, isometric) from the passive inactive condition as elevated bias: the CNN also captures correctly the pattern of EMG activity during combined conditions, however the scale of the estimate is substantially too low. The CNN correctly estimates isometric activity as increasing moment, positive joint rotation as decreasing moment, and alternating sign of joint moment during combined conditions. However again, the scale of the estimate is too low. The CNN captures the pattern of joint rotation and absence of rotation, but a scale that is too low.

Fig. 5 illustrates the ability of the CNN to extract simultaneous, independent changes in components of the neurobiomechanical state in both the superficial and the deep muscles. The CNN captures correctly the distinct, independent pattern and scale of EMG in GM and SO muscles. The CNN captures correctly the distinct, independent patterns and scales of joint rotation, activity and joint moment. While pattern and scale are correct in all quantities, there is a temporal error, decreasing through time, between ultrasound image derived estimates and the synchronised electrically recorded signals. This temporal error reduces the accuracy reported for these CNN estimates.

To summarise, the CNN separates the independent signals. During passive conditions the neural network was robust at predicting little to no active EMG, but a good proportion of passive motion. During isometric conditions the CNN predicted little to no passive motion, but a good proportion of active EMG. The reported accuracy is adversely affected mostly by errors in amplitude of prediction and also by some time varying temporal misalignment.

*C. Summary of Neural network performance*

For each muscle, and using just a single frame referenced to a common baseline frame, the neural network estimated meaningful values of all 3 signals (EMG, Joint moment, Joint angle). The mean absolute error (MAE) for Gas EMG, Sol EMG, Joint Moment and Joint Angle were 3.1±1.0 mV, 2.6±1.6 mV, 6.1±3.0 Nm and 2.5±1.3$^0$ (Table 1). In context, these errors represent 0.58, 0.56, 0.54 and 0.71 of the standard deviation of each signal, and 0.0056, 0.029, 0.056, 0.11 of the functional range of each signal where the ranges were 50 mV, 90 mV, 109 Nm and 23$^0$ respectively (Table 2). These results summarise 12,600 ± 2600 (mean ± S.D.) samples tested per participant and 403,023 samples in total, tested using cross-validation.

Accuracy, as a percentage of the time varying signal was 56.9±11%, 45.9±11%, 47.0±8.9%, 54.6±7.9% respectively (Table 1). Performance was consistent across all 32 participants with coefficients of variation in accuracy of 19%, 24%, 19% and 15% respectively (Table 1). There were no individuals for whom estimation of any signal failed (i.e. accuracy <0 or equivalently error larger than signal). Results for all 32 individuals are shown in Table A, Supp. Material.

Some very high individual accuracies were recorded for GM EMG with 3 participants at over 70%, and 13 at over 60%. There were only 4 participants with accuracy less than 40%. While accuracy was lowest for SO EMG, accuracy for several participants approached 70%. Estimation of ankle angle was the most stable with a high average accuracy of 55% and the lowest standard deviation at 7.9%.

## V. DISCUSSION

### A. The main finding

This investigation tests the hypothesis that three components (activity, length, tension) of the dynamic muscle state are encoded instantaneously and independently within the 3D-collagenous structure and are observable in generalizable form by 2D US images. We used novel data collection to generate 403,023 images from 32 participants containing independent and combined modulation of passive joint rotation and active neural input to the muscle state. We used deep convolutional neural networks to test whether the complex non-linear muscle state is encoded in 2D US images. Our results tested on 32 genuinely held out participants reveal that to approximately 50% accuracy (Table 1), the absolute values of EMG of each of the superficial and deep muscle, the ankle joint angle and the ankle joint moment is each encoded objectively, in generalizable form in 2D US images of the gastrocnemius and soleus calf muscles.

### B. Technical discussion

This manuscript demonstrates successful application of CNNs to predict continuous state variables rather than classify objects. This is US medical imaging analysis of physiological function rather than anatomical structure. A survey of literature in this journal (T-MI) indicates many papers applying CNN's to classification but few applying CNN's to regression. There is little published investigation of the architectures, hyper parameters and results of application of CNN's to regression and of the application of CNN's to modelling of complex systems such as skeletal muscle. Our use of DL to test a scientific hypothesis is novel. If the CNN can encode the neurobiomechanical state from US images of muscle, that demonstrates the state is encoded objectively within that muscle tissue.

This application is challenging and hence the neural network's ability to predict the absolute labels is surprisingly robust. The prediction of active and passive states from single images removes the possibility of signal drift. The CNN resolves active and passive states even when they vary independently and incoherently (c.f. Fig. 5, EMG (GM) and JA at 20-45s). The encoding of this information in the US image is not trivial or intuitive. To illustrate the achievement, observe the motion of the muscle in Video 1. Supplementary Material, and note that from a *single frame*, and a *reference frame*, the active and passive states of each muscle are estimated to ~50% accuracy.

The necessity of the reference frame is to reduce bias and promote generalisation between participants. We anticipate a reference frame would not be necessary to generalise to new motions within the same participant. Within participant generalisation is relevant to prosthetics, where a system is trained on person-specific actions to control a prosthetic limb [33], [34]. Our existing evidence suggests a within participant system would be more accurate than a general system [73].

Several sources of error limit the accuracy of this generalised system to approximately 50% (Table 1).

(i) Temporal misalignment between US images and measured signals occurs and is variable (Fig 5, c.f. 8-10s, and 40-45s). EMG and joint moment signals contain more or their power at higher frequencies and are thus more sensitive to temporal misalignment. Substantial temporal misalignment is a known phenomenon related to image timing in clinical US interfaces [74]. While clinical US interfaces such as the one we used, give no control over image timing, accurate image timing is possible using low-level US systems [74]. For this study, misalignment means the reported accuracy is less than the intrinsic accuracy of the CNN estimate (c.f. Fig 5).

(ii) Un-encoded variation between participants: Contrary to our hypothesis, some variation between participants is intrinsically not encoded within the US images. Variation between participants in limb strength, muscle mass, muscle cross sectional area, electrical electrode-skin-muscle impedance, electrode placement, location of the foot the footplate, is not available to the CNN. These variables alter the mapping between US image and measured signals (EMG, joint angle, joint moment). Our results provide the first benchmark of the generalizable content of US images. Injecting some prior knowledge into the system such anthropometric data would likely improve accuracy. Within participant generalisation would avoid these issues and show higher accuracy.

(iii) Imperfect ground truth: The measured signals have limited accuracy. Imperfect placement of EMG electrodes, partial sampling of the whole muscle volume, crosstalk from adjacent muscles, electrical noise and interference, foot placement, slight knee flexion, heel raising and toe curling limit accuracy of EMG, joint angle and joint moment signals.

(iv) US probe placement: The US acquires a single plane from a 3D muscle structure. The extent to which generalisation depends upon the specific plane imaged is uncertain. More data is required to test the effect of variation in probe placement.

Given the uncertainties listed above, it remains remarkable that from a single 2D US image referred to a baseline image, absolute values of each component of the neurobiomechanical state can be estimated on new participants to a benchmark accuracy of ~50%.

Our empirical demonstration provides the first proof of principle that estimation of specific muscle states in deep muscle is possible in general conditions of combined/isolated active and passive changes. The deep soleus muscle (SO) gave EMG accuracy comparable within 10%, to the superficial GM. While we placed the probe to acquire GM and SO within a single image, one plane is not optimal for both muscles. We chose to optimise for GM. Thus our benchmark for SO represents a lower limit to what is possible for this muscle.

### C. Application to other muscles

In this investigation we chose muscles for their suitability to test our hypothesis that muscle states are encoded in their collagen structure and are observable by US. This investigation required the muscles to be observable by US and required the main inputs of EMG and passive joint rotation to be measureable. This investigation also required it to be possible to control the two inputs experimentally to produce a data set covering the space of single and simultaneous combined, independent variation of neural activity and passive joint rotation. The calf muscles are a relatively well understood muscle group with access to control and measure these labels.

However, the significance of this investigation lies in potential to measure neural output more generally from deep muscles. Significance lies also in potential to measure tension generally from individual muscles rather than joint moments. Here the CNN was trained to predict joint moment, which is measureable. However the prediction was derived from an image of muscle which means that information, in essence muscle tension, is encoded within the muscle.

Following our test of principle, the question becomes how to acquire the labels and training data to train a system on more general muscle groups? In principle, though with greater practical difficulty, it is possible to record EMG with needles/wires from deep muscles inaccessible via surface electrodes. This could give training labels for muscle in a complex system like the neck, back or forearm. It is also possible to use unsupervised learning (like Bayesian GAN [75]) on a large data collection of US only, and reserve supervised learning for a smaller data collection of US with EMG labels. Another possibility is to measure dynamometric or kinematics signals (e.g. head torque), and predict those signals directly from the image, in such a way that the network learns a spatial localisation mapping from the labels to the image (like class activation mapping; CAM [76]). Combined with an accurate segmentation [7], [77], an activity map of generated head force, could provide estimated muscle-specific contribution to gross head rotational force.

### D. Scientific and clinical significance

We present the first generalized prediction of independent components of the neurobiomechanical state of skeletal muscle (activity, joint rotation, joint moment) directly from standard frame-rate (25Hz) 2-D, b-mode US in general conditions of independently varying inputs (Figs. 4-5). This result reports a scientific discovery. Previously, it was unknown whether muscle tissue encodes, simultaneously, the activity, origin-insertion length and tension state of the muscle.

This scientific discovery has technological significance. *Muscle activity* is an amplified version of neural output or motor command delivered by peripheral nerves from the spinal cord to the muscle [78]. Non-invasive measurement of activity in deep muscles is currently impossible. In science, we need activity, particularly from deep muscles in the neck, back, lower and upper limbs, to understand how control of the muscular system is organised. Muscular control is hierarchical and synergistic in nature, and currently that science is immature simply because we cannot measure activity easily in all the important deep muscles [6]. In medicine, healthy control of muscles has broken down for many possible reasons. In myopathies or injury the muscle is inflamed or diseased and delivers inadequate output from neural drive. In neuropathies, neural drive to the muscle is inadequate e.g. from breaks in the peripheral nerves or spinal cord injury, or from demyelination of upper (central) or lower (peripheral) motor neuron. In neurological conditions control of muscles is disordered e.g. dystonia, Parkinson's, cerebella ataxia resulting in abnormal patterns and timing of activity. Measurement of activity can discriminate myopathies from neuropathies from neurological conditions and localise the impairment. In the clinic, staff rely routinely on manual palpation and very rarely on needle EMG since that is an expert skill in very short supply. Our US approach, developed to its potential, could provide easy discrimination between these conditions and diagnosis of abnormal muscle activity. In rehabilitation, prostheses are controlled where possible using available activity signals from muscles. While it is possible to control prostheses directly from brain interfaces, peripheral neural output provides better quality signals since these are already pre-processed motor signals, and muscle activity is simply an amplified version of peripheral neural output [78]. There is current interest in using wearable US to drive prosthetic devices [33], [34].

*Muscle tension*: Measurement of individual muscle force requires a strain gauge inserted surgically in series with the individual muscle, typically in the tendon joining that muscle to bone [79]. For surgery, orthopaedics, rehabilitation, and biomechanics the force of individual muscles is needed to determine the contribution individual muscles make to joint moments and joint stability, to whether the balance between muscles is correct and whether surgical correction, physiotherapy or altered training is required. In this study, we have validated the CNN estimation of force using a joint moment. However, the estimate was derived from muscle tissue, and thus our results demonstrate that force can be estimated directly from muscle tissue. This study provides objective evidence to justify surgical implantation of strain gauges to provide muscle specific force labels.

*Muscle stiffness*: Currently it is believed that observational US can only measure muscle strain, and because stress is unknown, cannot measure mechanical properties such as force or stiffness. The result published here reverses that belief and reveals that analysis of the full biomechanical state (length and tension) is possible using observational imaging. Muscle force and stiffness are a consequence of intrinsic muscle properties operating on the inputs (neural command and joint rotation). The proof of concept demonstrated in this paper is that multilayered neural networks with DL methods (convolutions, pooling, dropout, etc…) can model directly from US images the intrinsic muscle properties, and the independent inputs which together determine the mechanical output. This result is possible because the collagenous structure of skeletal muscle is observable, and also because muscle activity and passive joint rotation create different patterns of strain within the structure [9]. Force generated internally by activity within individual motor units has a different strain pattern to force transmitted externally into the muscle between origin and insertion. Bypassing human preconception, ANN's can learn those dynamic non-linear patterns and provide spatiotemporal representations of the muscle state for our scientific and diagnostic benefit.

These results imply that perturbation methods (e.g. shear wave Elastography) are not required measure the biomechanical state. In practice, further development will be required to translate this proof of principle into a technology applicable to all muscles of medical interest. Standard US machines are more available, more cheaply than shear wave imaging machines and they input less acoustic power to the patient. For dynamic structures as complex as skeletal muscle, data driven modelling of muscle properties using DL should be more accurate than using generic stress -strain relationships and assumptions of material properties to interpret shear wave velocity maps.

### VI. CONCLUSIONS

Currently, there is an unmet need for technology to provide non-invasive diagnosis of skeletal muscle state in general conditions. Limitations in currently technology (electromyography, dynamometry, shear wave elastography) mean that many important muscles (e.g. deep muscles in the neck, back, thorax/abdomen and limbs) are inaccessible to full diagnostic analysis. This paper demonstrates an approach which can contribute new diagnosis of the muscle system.

We have presented a novel experiment for the generation of hundreds of thousands of accurately labelled muscle US images for modelling functional muscle states using US. We have demonstrated that skeletal muscle encodes three components of the neurobiomechanical state within its tissue structure, observable by US. We have presented the first generalized prediction of muscle specific EMG, of joint angle and joint moment from standard frame-rate b-mode, 2D US images. Existing methods rely on simple measures in isolated cases (isometric only, or passive only) which do not generalize. We have demonstrated the efficacy of CNNs to this domain, which encourages application of deep learning to skeletal muscle US. This approach has potential applications to clinical diagnosis, monitoring of treatment, biofeedback for behavioural therapy and interfacing with prosthetics in large range of conditions of substantial socioeconomic impact as stated in the introduction.

## VII. Acknowledgements

We thank all the participants who gave their time generously. With appreciation we thank Des Richards for his work building the bespoke footplate apparatus for our experiment.